\documentclass[pra,twocolumn]{revtex4}
\usepackage{amssymb,amsmath,epsfig,bm,color,slashed,mathrsfs}
\newcommand{\be}{\begin{equation}}
\newcommand{\ee}{\end{equation}}
\newcommand{\bea}{\begin{eqnarray}}
\newcommand{\eea}{\end{eqnarray}}

\newcommand{\vecu}{\bm u}
\newcommand{\veczeta}{\bm \zeta}
\newcommand{\vecOm}{\bm \Omega}
\newcommand{\vecnabla}{\bm \nabla}

\newcommand{\vecx}{\bm x}

\newcommand{\ie}{{\it i.e., }}
\newcommand{\eg}{{\it e.g., }}

\definecolor{red}{rgb}{0.8,0,0}
\definecolor{RED}{rgb}{0.8,0,0}
\definecolor{violet}{rgb}{0.4,0,0.4}
\definecolor{green}{rgb}{0,0.5,0.0}
\definecolor{GREEN}{rgb}{0,0.5,0.0}
\definecolor{navy}{rgb}{0.0,0.0,0.6}
\definecolor{orange}{rgb}{0.8,0.2,0.0}
\definecolor{blue}{rgb}{0.3,0.0,0.8}

\begin{document}

\title{Spontaneous  symmetry breaking in 
 rotating condensates of ultracold atoms}
   
\author{Armen Sedrakian}
\affiliation{
Institute for Theoretical Physics, 
J.~W.~Goethe-University, \\D-60438  Frankfurt-Main, Germany
}

\begin{abstract}
  We describe an equilibrium state of a rotating trapped atomic
  condensate,  which 
  is characterized by a non-zero internal circulation and spontaneous
  breaking of the rotational $O(2)$ symmetry with all three major
  semiaxes of the condensate having different values. The
  macroscopic rotation of the condensate is supported by a mesh of
  quantized vortices, whose number density is a function of internal
  circulation.  The oscillation modes of this state are computed
  and the Goldstone mode associated with the loss of the symmetry is
  identified. The possible avenues for experimental identification
  this state are discussed.
\end{abstract}

\maketitle

{\it Introduction.}
Rotating condensates of ultracold Bose or Fermi
gases form vortex lattices above a critical rotation frequency. Such
states have indeed been observed in a number of experiments on superfluid
bosonic~\cite{VorticesBosons} and fermionic~\cite{VorticesFermions}
vapors confined to magnetic and/or optical traps. These systems
provide an excellent tool to study various aspects of ultracold
rotating atomic gases due to the high level of control over the parameters,
such as the number of vortices, strength of inter-particle
interactions, shape and rotation rate of the trapping potential,
multi-species loading of the trap, etc.  Various aspects of the
rotating condensates, have spawned an enormous literature (see, \eg
\cite{Fetter,Giorgini}).  For our purposes it should be pointed out
that theoretical studies of the rotational equilibrium of
condensates featuring an array of vortices and their oscillations have
confirmed the applicability of ``coarse-grained'' superfluid
hydrodynamics for such
systems~\cite{SW01,Chevy:2003,2003PhRvA..68c1605C,2004EL.....65..594S,2006PhRvA..73a3616W,2006PhRvA..73d1604C,2007PhRvA..76c1601W,2007PhRvA..75e3609A,2009PhRvA..80b3610L,2012JLTP..166...49C}.
In particular, an excellent agreement between the theory and
experiment is observed for the breathing modes of a uniformly rotating
condensate~\cite{2006PhRvA..73a3616W}.  The applicability of 
superfluid hydrodynamics needs a sufficiently large and strongly
coupled condensate, which can be routinely created in experiments.
Under these conditions and in the case of a Bose gas, the quantum
pressure term in the Gross-Pitaevskii equations can be ignored
compared with the interaction terms, in which case this theory takes
the form of superfluid hydrodynamics (see Ref.~\cite{SW01}). 
Furthermore, the size of the vortex core, which scales
with the number of particles as $\xi \sim (8\pi a N)^{-1/2}$, where
$a$ is the scattering length, is sufficiently
small in this case and need not be resolved. 

In this work we explore the rotational state and oscillations of {\it
  non-uniformly rotating} condensates. We concentrate on a special
class of departures from rigid body rotation, which feature a constant
condensate circulation in the frame rotating with the surface of the
condensate.  The states that we are seeking are supported by
a sufficiently dense mesh of quantum vortices, which guarantee 
non-zero circulation in the laboratory frame. Bifurcation in
condensates at overcritical rotation rates was considered previously
with the constraint of irrotationality of the superfluid
velocity~\cite{Recati}. Although such states do spontaneously break the rotational
symmetry, they are generically unstable and, thus, their time-dependent
dynamics is directed towards  formation of vortices in the condensate.

We formulate the superfluid hydrodynamics in its virial
form, which was proposed and applied previously to uniformly rotating
condensates in Refs.~\cite{SW01}. The method has its roots in 
developments in the context of self-gravitating
fluids~\cite{Chandrasekhar69}. There are some parallels (duality),
but also differences between gravitationally bound and trapped
systems~\cite{SW01}.  The states of the condensates discussed below
are the duals to the gravitationally bound {\it Riemann ellipsoids}.

{\it Equilibrium.}
Consider a rotating cloud of condensed gas
confined by a harmonic trapping potential $\phi_{\rm tr} (\vecx) =
m\omega_i^2x_i^2/2$, where $m$ is the atom mass and $\omega_i$ are the
Cartesian trapping frequency components (hereafter the Latin indices
run through 1, 2, 3 -- the components of the Cartesian coordinate
system --
and a summation over the repeated indices from 1 to 3 is assumed).
Below we consider the case of axisymmetric trapping with $\omega_1 =
\omega_2 \equiv\omega_{\perp}$.  We will assume below, that the
rotation frequencies are well above the lower critical frequency for
nucleation of the vortices, which scales as $\Omega_{c1}\propto
(\hbar/mR^2) \ln (R/\xi)$, where $R$ is the size of the condensate
transverse to the rotation. Then, the condensate executes rigid-body
rotation, where its angular momentum is supported by a vortex lattice.
Hydrodynamical treatments of such systems are carried out after
averaging (``coarse graining'') the dynamical quantities over
distances much larger than the inter-vortex distance.  The rotating
condensate is then described by the Euler equation, which we write in
a frame rotating with some angular velocity $\vecOm$ \bea\label{EULER}
\rho\left( \partial_t +u_{j}{\nabla_j}\right) u_{i}
&=&-\nabla_i p -\frac{\rho}{2}\nabla_i (\omega_j^2x_j^2)\nonumber\\
&+&\frac{\rho}{2}\nabla_i\vert\vecOm\times\vecx\vert^2
+2\rho\epsilon_{ilm}u_{l}\Omega_m,
\label{eq:euler}
\eea where $\rho$, $p$, and $u_i$ are the density, pressure, and
velocity of the condensate. We will specify the rotating frame more
precisely later on.  Equation \eqref{EULER} is valid
for  Bose or Fermi gases, provided the appropriate equations of state are used.
  The first moment of Eq. \eqref{eq:euler},
integrated over the condensate volume $V$, is the second-order tensor
virial equation \bea\label{second_order_virial} \frac{d}{dt}\int_{V
}{d^3x\rho x_j u_{i}} &=&2\mathscr{T}_{ij}+\delta_{ij}\Pi
+(\Omega^2 - \omega_i^2) I_{ij}\nonumber\\
&&\hspace{-1cm}+2\epsilon_{ilm}\Omega_m \int_{V }{d^3x\rho x_ju_{ l}}
-\Omega_i\Omega_kI_{kj},
\label{eq:v2}
\eea where 
$ I_{ij}\equiv \int_V{d^3x\,\rho x_ix_j}$ and  $\mathscr{
  T}_{ij}\equiv (1/2)\int_{V }{d^3x\rho u_{ i}u_{ j}}, 
$
are the second-rank tensors of the moment of inertia and kinetic energy,
and $\Pi= \int_V d^3x p$ is the scalar volume integral of the
pressure.  A direct integration of the unperturbed hydrodynamic
equation for the condensate for the class of polytropic equations of
state shows that the density distribution is of a Thomas-Fermi type, \ie
the density is a quadratic form of the coordinates $\rho = \rho(l^2)$,
where $ l^2 ={x_1^2}/{a_1^2}+{x_2^2}/{a_2^2}+{x_3^2}/{a_3^2}$~\cite{SW01}. For
such a distribution the moment of inertia tensor can be written as $
I_{ij} = ({4\pi}/{3})a_i^3a_ka_l\delta_{ij} \int_0^1\rho(l^2) l^4dl, $
which will allow us to translate the equations written for moment of
inertia tensors to equations written for the semiaxes of the equilibrium
condensate.

The equilibrium shape of the condensate in a rotating axisymmetric
trapping potential is an ellipsoid of revolution.  We fix the origin
of the {\it rotating coordinate system} to coincide with that of the
ellipsoid, while the axes of the system coincide with the semiaxes $a_1$,
$a_2$ and $a_3$ of the ellipsoid. Thus, the system of coordinates
co-rotates with the axis of ellipsoid at a frequency $\vecOm = (0, 0,
\Omega)$ introduced above. We now specialize our discussion to a class
of solutions for which $\vecnabla\times \vecu = \veczeta,$ \ie
solutions which admit {\it non-zero vortical motion of the fluid in
  the rotating frame}.  We consider the simplest case of uniform
vortical motion  along the rotation vector, in which case $ u_1 =
Q_1 x_2, \quad u_2 = Q_2 x_1, \quad u_3 = 0, $ where $Q_1$ and $Q_2$
are constants such that $\zeta_3= Q_2 - Q_1 \equiv \zeta $. We further
assume that the fluid flows associated with internal motions are
non-advective, \ie satisfy the condition $\vecu\cdot 
\vecnabla\rho = 0, $ which is manifestly the case for,
\eg incompressible fluids. This implies $ {u_1x_1}/{a_1^2} =
-{u_2x_2}/{a_2^2}$ or $Q_1/a_1^2 = -Q_2/a_2^2$, 
$ Q_1 = -a_1^2\zeta/(a_1^2+a_2^2)$ and $Q_2 =
a_2^2\zeta/(a_1^2+a_2^2)$.

Consider first the stationary solutions of
Eq.~(\ref{second_order_virial}).  We write down the diagonal elements
of this equation and  eliminate $\Pi$, to obtain \bea -\omega_3^2I_{33}
&=& 2\mathscr{T}_{11} + (\Omega^2-\omega_\perp^2) I_{11} + 2\Omega
\int\!\! d^3x \rho x_1 u_2\nonumber\\
&=& 2\mathscr{T}_{22} + (\Omega^2-\omega_\perp^2) I_{22} - 2\Omega
\int\!\!  d^3x \rho x_2 u_1 .  \eea Substituting the components of the
velocity $u_i$ and after some further manipulations one finds 
\be\label{eq:triangle}
a_1^2 q_{\perp}^2 = a_2^2 q_{\perp}^2  = \omega_3^2a_3^2 +2\Omega
\zeta {a_1^2a_2^2}/{(a_1^2+a_2^2)},
\ee 
where $q_{\perp}^2 \equiv Q_1Q_2 +\omega_{\perp}^2  -\Omega^2 $,
These
equalities are satisfied if $\Omega^2-Q_1Q_2 = \omega_\perp^2$ and
$-2a_1^2a_2^2 \zeta\Omega/(a_1^2+a_2^2) = \omega_3^2a_3^2.$ Note that
these relations can be equivalently obtained from a variational
principle, where the energy of the system is minimized with respect to
the parameters $a_2$ and $a_3$ at fixed mass, angular momentum and
circulation of the fluid~\cite{Lai:1993ve}.  It is useful to define
the ratio $ f\equiv {\zeta}/{\Omega}$, in terms of which these
equilibrium conditions read
\bea \label{eq:struc} \bar\Omega^2
\left[1+\frac{\delta_2^2}{(1+\delta_2^2)^2}f^2\right] = 1, \quad
-\frac{2\delta_2^2}{1+\delta_2^2}f \bar\Omega^2 =
\bar\omega^2\delta_3^2, 
\eea 
where the reduced quantities are defined
as $\bar\Omega = \Omega/\omega_{\perp}$, $\bar\omega =
\omega_3/\omega_{\perp}$ with $\omega_{\perp}$  the transverse
trapping frequency,  $\delta_{i} = a_{i}/a_1$, $i = 2, 3$, and all
the lengths are measured in units of $a_1 = 1$.
\begin{figure}[tb] 
\begin{center}
\includegraphics[width=7cm,height=6cm]{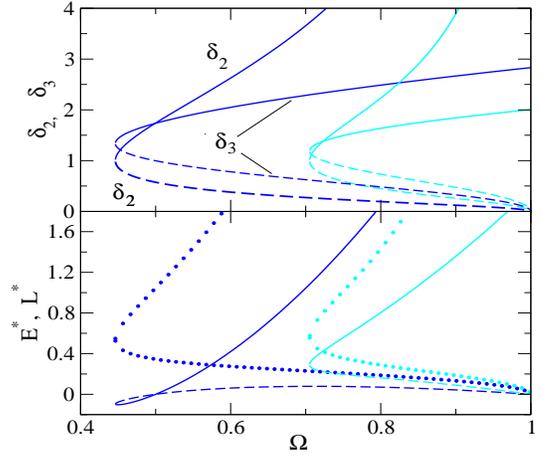}
\caption{ (Color online)
Upper panel: The dimensionless axis $\delta_2$ 
and $\delta_3$  as  functions of the normalized rotation frequency
for fixed $f= 4$ (dark, blue online) and $f=2$ (light, cyan lines). 
The solid and dashed lines distinguish the configurations.
Lower panel: The energies of the first (solid lines) and second
(dashed lines) configuration for $f=2$ and $f=4.$ 
Their  angular momenta are shown by dotted lines.
}
\label{fig:1}
\end{center}
\end{figure}
Equations \eqref{eq:struc} fully determine the equilibrium structure
of the cloud for any pair of values of $\zeta$ and $\Omega$.  Note
that the first equation implies that the rotation frequency must be
{\it below} the transverse trapping frequency, \ie $\omega_{\perp} >
\Omega$, as is the case also for uniformly rotating condensates.  The
solutions (\ref{eq:struc}) represent an example of spontaneous
symmetry breaking (SSB) in rotating condensed clouds: the initial
Lagrangian and the associated hydrodynamical equations have manifest
$O(2)$ symmetry, but the ground state does not. Thus, there exists a
minimal frequency above which $a_1\neq a_2\neq a_3$ as a consequence of
$\zeta\neq 0$.  Indeed, in the limit $\zeta \to 0$
Eqs.~\eqref{eq:triangle} give $(\bar \Omega^2 -1) =\delta_2^2(\bar
\Omega^2 -1) = -\bar \omega^2\delta_3^2$, which implies that
$\delta_2^2 = 1,$ $\delta_3^2 = {(1-\bar\Omega^2)}/{\bar\omega^2}$, \ie
if $\zeta = 0$ the rotational $O(2)$ symmetry is unbroken. The
explicit solutions of Eqs.~(\ref{eq:struc}) read $\delta_2^2 = (c-1)
\pm \sqrt{(c-1)^2-1},$ where $c = f^{2}/2 (\bar\Omega^{-2}-1)$ and
$\delta_3^2 = -{2f \bar\Omega^2\delta_2^2
}/(1+\delta_2^2)\bar\omega^2.$ Since the radical needs to be positive
in order for $\delta_2^2$ to be real, we conclude that there exists a
minimum rotation frequency 
\be \label{OmMin}
\bar \Omega \ge \bar\Omega_{\rm min} =
\left(1+{f^2}/{4}\right)^{-1/2},
\ee 
above which the SSB sets in. 

The energy and angular momentum of the condensate 
cloud associated with the solutions above are given by
\bea
E^*  =\frac{E}{\omega_{\perp}^2} 
&=& \frac{1}{2} 
\left[- (1+\delta_2^2)-\bar\omega^2\delta_3^2\right],
\nonumber\\
&+& \frac{1}{2} (1+\delta_2^2) (1+b^2) \bar\Omega^2 
  - 2 \delta_2 b\bar\Omega^2\\
L^* = \frac{L}{M\omega_{\perp}} &=& \frac{1}{5}\left[
(1+\delta_2^2) - 2\delta_2b\right]\bar\Omega,
\eea
where $b = -f\delta_2/(1+\delta_2^2)$ and $M$ is the mass of the cloud.

\begin{figure}[tb] 
\begin{center}
\includegraphics[width=10cm]{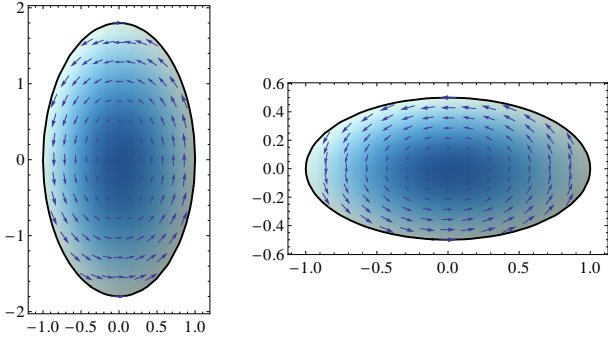}
\caption{ Illustration of the velocity vector field in the $x$-$y$ plane
  for two possible solutions with SSB for $\Omega = 0.5$, $f=4$ and
  $\delta_2=1.8$ (left figure) and  $\delta_2=0.5$ (right figure). Note
  that the figures are not to the same scale. The third axis values are
  $\delta_3 = 1.5$ and $0.98$ respectively. The color coding reflects
  the Thomas-Fermi density distribution of the cloud.}
\label{fig:2}
\end{center}
\end{figure}

Figure~\ref{fig:1} shows the values of the
(dimensionless) semiaxes of the condensate; for $\Omega
<\Omega_{\rm min}$ only axisymmetric figures exist with $a_1 = a_2 \neq
a_3$; for $\Omega \ge \Omega_{\rm min}$ two pairs of non-axisymmetric
solutions exist, which are shown by solid and dashed lines. The figure also
shows the energy and angular momentum of these states. 
It is seen that the solutions 
with small ($\delta_2, \delta_3 <1$) semiaxes have lower energy
than the ones with large  ($\delta_2, \delta_3 >1$) semiaxes in units
of $a_1$. Close to the onset of axial nonsymmetry the energies of both
configurations are nearly degenerate, as expected. 
The velocity fields for two possible solutions for fixed values of rotation
frequency $\Omega = 0.5$ and circulation $\zeta = 2$ are shown in
Fig.~\ref{fig:2}.

{\it Oscillation modes}.
Consider small perturbations around the equilibrium with time
dependence $\exp (\lambda t)$. The lowest-order non-trivial
perturbations are those with $l=2$ and $\vert m\vert \le 2$.  The
Lagrangian perturbation of Eq.~\eqref{second_order_virial} is given by
\bea &&\lambda^2 V_{i:j} - 2 \lambda Q_{jl} V_{i;l} - 2\lambda \Omega
\epsilon_{il3}
V_{l;j} \nonumber\\
&&\hspace{1cm}- 2\Omega\epsilon_{il3} (Q_{lk} V_{j;k}-Q_{jk} V_{l;k})
+ Q^2_{jl} V_{i;l}+ Q^2_{il} V_{j;l}\nonumber\\
&&\hspace{1cm}=\Omega^2 (V_{ij} -\delta_{i3}V_{3j}) -\omega_i^2 V_{ij}
+\delta_{ij}\delta\Pi, \eea where $V_{i;j} = \int_V d^3x\xi_i x_j$
with $\xi_i$ being the Lagrangian perturbation, and $V_{ij} =
V_{i;j}+V_{j;i}$.  In the case when the internal circulation $\veczeta$ is
along the spin vector, the only nonzero elements of the matrices $Q$
and $Q^2$ are $Q_{12} = Q_1, Q_{21} = Q_2$ and $Q^2_{11} = Q_1Q_2,
Q_{22}^2 = Q_1Q_2$. The modes that have even and odd parity with
respect to the index 3 decouple.  To determine the even modes we need
the equation of state of the condensate, which determines the pressure
perturbations. For the class of fluids described by a polytropic
equation of state $p = \rho^{\gamma}$ the pressure perturbation is
given by $\delta \Pi = (1-\gamma)\left[
 (\omega_\perp^2-\Omega^2)(V_{11}+V_{22}) +\omega_3^2V_{33}\right]/2$~\cite{SW01}.  
In the case of an incompressible condensate one can proceed in a model-
and statistics-independent way, because in that case the perturbations are
solenoidal, which translates into $ V_{11}a_1^{-2}+
V_{22}a_2^{-2}+V_{33}a_3^{-2} = 0.$ The characteristic equation for
even modes in that case is
\bea {\rm Det} \left(
\begin{array}{ccc}
\frac{\lambda^2}{2}+p^2- \Omega Q_2
&  -3\Omega  Q_1
&  - \frac{\lambda^2}{2}-\omega_3^2\\
+3\Omega  Q_2
&\frac{\lambda^2}{2}+p^2+ \Omega Q_1
& - \frac{\lambda^2}{2}- \omega_3^2\\
a_1^{-2} &a_2^{-2} &a_3^{-2} \\
\end{array}\right)  = 0,\nonumber
\eea with the abbreviation $p^2 = \Omega^2+\omega_{\perp}^2-Q_1Q_2$.
In the compressible case the elements in the last row should be
replaced: $a_1^{-2}\to (\gamma-1)(\omega_\perp^2-\Omega^2)/2$, the
same for $a_2^{-2}$, and finally $a_3^{-2} \to
\lambda^2/2+(1+\gamma)\omega_3^2/2$.  Note that the characteristic
equation for the modes is invariant under simultaneous interchange
$Q_1 \leftrightarrow -Q_2$. Below we assume
a Bose gas with $\gamma = 2$. The characteristic equation for even modes
is of order 8 in the incompressible case and order 12 in the
compressible case. However, the modes come in complex conjugate pairs
and the number of distinct modes is reduced.  The odd-parity modes are
given by \bea {\rm Det} \left(
\begin{array}{ccccc}
  \lambda^2  +\tilde \omega_{\perp}^2 
  &- 2\lambda \Omega &
  q_{\perp}^2 -2\Omega Q_2 & 0\\
  2\lambda \Omega & 
  \lambda^2 +\tilde\omega_{\perp}^2 
  &0&
  q_{\perp}^2+2\Omega Q_1\\
  \omega_3^2 & 0 &\lambda^2  + q_3^2
  & - 2\lambda Q_1 \\
  0 & \omega_3^2 &- 2\lambda Q_2& 
  \lambda^2  + q_3^2 
\end{array}\right) = 0.\nonumber
\eea
with the abbreviation $q_3^2 = Q_1Q_2 + \omega_3^2$.

\begin{figure}[tb] 
\begin{center}
\includegraphics[width=7.0cm,height=8.0cm]{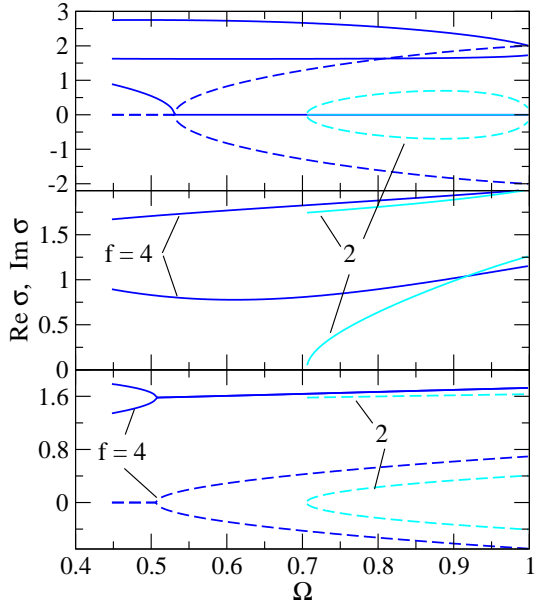}
\caption{ The even-parity compressible (upper panel) and
 incompressible (middle panel) modes along with the odd-parity (lower
 panel) modes for $f=2$ (light, cyan online), and $ 4$ (dark, blue
 online) as a function of rotation frequency. All quantities are
 normalized by the transverse trapping frequency
 $\omega_{\perp}$. The real parts are shown by the solid lines, the
 imaginary ones by the dashed lines.}
\label{fig:3}
\end{center}
\end{figure}

The secular equations for the even and odd  modes were
solved numerically and their real and imaginary parts are shown in
terms of the quantity $\sigma = -i\lambda$ in Fig. \ref{fig:3}. 
The real parts of the $\sigma$'s are the eigen-frequencies of
the modes and their imaginary parts describe their  damping. 
Although for each pair of values of $\Omega$ and $f$ there are two equilibrium
background solutions which exhibit SSB, there is only one set of
oscillation modes associated with both solutions for $\delta_2$
and $\delta_3$. This is consequence of the  $Q_1\leftrightarrow -Q_2$
exchange invariance mentioned above.

There is a trivial undamped odd-parity mode $\sigma_1 = \Omega$, which
arises because we are working in the rotating frame.  A second
undamped mode with eigen-frequency $\sigma_2 = \sqrt{- Q_1Q_2}$ can be
identified with the Goldstone mode, which emerges as a consequence of
the SSB; indeed $\sigma_2\propto \zeta$ and vanishes in the limit
where the internal circulation is zero.  The remaining odd-parity
modes are shown in Fig.~\ref{fig:3}. There are two physically distinct
domains: in the low-rotation-frequency domain there are two real modes
without damping; in the high-rotation-frequency domain the real parts
are degenerate, whereas the imaginary parts are equal and opposite in
sign. The negative imaginary part, emerging in this domain, indicates
{\it dynamical} (\ie non-dissipative) instability of the system
towards even parity oscillations. The two domains are separated by the
point of onset of dynamical instability. Furthermore, we see that
the stable segment decreases with decreasing $f$ and is absent for
$f\le 2$ in our example.
There are two real distinct even modes of oscillations in the
incompressible case, which are shown in the middle panel of
Fig. \ref{fig:3} for $f = 2,$ and $ 4$ as functions of rotation
frequency. The high-frequency mode weakly depends on the value of $f$,
whereas the low-frequency mode changes its asymptotic behavior in the
low-rotation-frequency limit. The incompressible even modes are purely real and
correspond to undamped oscillations. The compressible even modes,
displayed in the upper panel of Fig. \ref{fig:3} show dynamical
instability for high rotation rates, whereby a stable branch of
oscillations appears for $f\le 3$ and low rotation rates.  

\begin{figure}[tb] 
\begin{center}
\includegraphics[width=7.5cm,height=6.5cm]{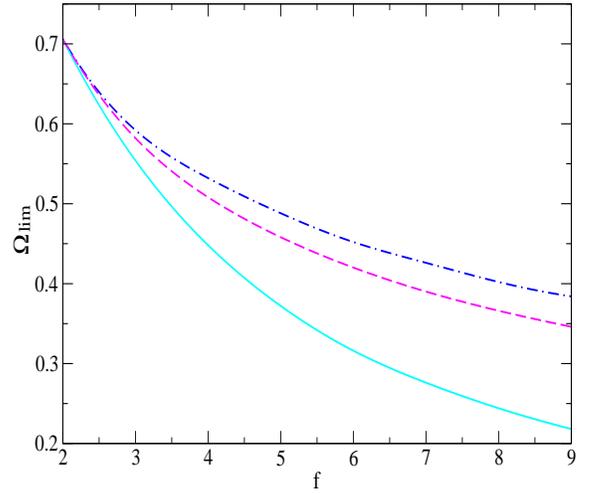}
\caption{ Dependence of the limiting dimensionless rotation rates
  $\Omega_{\rm lim}$ for onset of SSB, Eq. (\ref{OmMin}),
  (above the solid line), for the onset of instability of
  compressible even modes (above dashed line) and instability of odd
  modes (above dash-dotted line) on the internal circulation parameter
  $f$. The region corresponding to stable SSB state is enclosed
  between the solid and the dashed lines.
}
\label{fig:4}
\end{center}
\end{figure}

Now we locate the  space spanned by the parameters $\bar\Omega$
and $f$, where the SSB state is both energetically favorable and stable
against the odd- and even-parity oscillations. In Fig.~\ref{fig:4} the
region above the solid line, which shows the $\Omega_{\rm lim}(f)$
dependence given by Eq.~(\ref{OmMin}), features the SSB state. This
region is bound from above by the minimal frequency at which the even
compressible oscillation modes become unstable. Thus, the region
between the solid and dashed lines in  Fig.~\ref{fig:4} corresponds to
the parameter space where the SSB state is stable. It is further seen that the minimum
frequency at which the odd-parity modes become unstable is larger than
that for the even-parity modes for all values of $f$, \ie, the
stability of the SSB state is controlled by the even parity modes alone.

{\it Experimental verification}.
The state of rotating condensates with internal
circulation can be studied with experimental set-ups
already used for uniformly rotating clouds. First, one needs to
establish the SSB by, \eg measurements of the axis ratio via
imaging the condensate. The knowledge of the
axis-ratios will permit reconstruction of all other parameters of the
cloud, including the magnitude of the internal circulation. Second,
the number of vortex lines in the cloud can be ``counted'' again
through imaging. Since the equilibrium rotation of the condensate can
be independently measured through the excitation of the surface modes,
one can detect potential deviations from the Feynman formula $n_v =
2\Omega/\kappa$, where $n_v$ is the number density of vortices and
$\kappa$ is the quantum of circulation.  A breakdown of the Feynman
formula would be evidence for internal circulations.
Indeed, if internal circulations are present then the vorticity in the
laboratory frame and the vortex density are given respectively  by 
\be \label{eq:generalized_feynman}
\zeta_L = 
(2 + f) \Omega, \quad n_v = (2+f)\frac{\Omega}{\kappa}.  
\ee Thus, a
simultaneous and independent measurement of $\Omega$ and $n_v$ will
give the experimental value of $f$. Third, the oscillation modes
can be measured and tested against theoretical predictions, as has already
been done for the $l=2$ modes of uniformly rotating condensates.
 
{\it Conclusions.}
In this work we identified a state of a rotating, harmonically
trapped, condensate of an atomic cloud, which in the frame rotating
with the cloud's surface has non-zero internal circulation. The resulting
equilibrium configurations are non-axisymmetric, and thus are a
manifestation of SSB in superfluid hydrodynamics. We have derived
the complete set of $l=2$ harmonic modes, which are dynamically
unstable for high rotation rates and low internal circulation and are
stable otherwise. Several experimental tests have been suggested,
which can shed light on the structure and small-amplitude oscillations
of non-uniformly rotating clouds of Bose and Fermi condensates.  

As studies of rotating condensates provide a useful test bed for
exploring strongly correlated systems under rotation, due to the great
experimental flexibility available in these systems, further insights
are expected about systems that are difficult or impossible to
manipulate and/or observe. One such example is provided by rotating neutron
stars containing strongly interacting condensates of nuclear and quark
matter.

I am grateful to S. Stringari, H. Warringa, and M. Zwierlein for
discussions.  I also acknowledge the Aspen Center for Physics, where
the major part of this work was done.

\end{document}